# Potential Fluctuations at Low Temperatures in Mesoscopic-Scale $SmTiO_3$/$SrTiO_3$/$SmTiO_3$ Quantum Well Structures

*Will J. Hardy[1], Brandon Isaac[2], Patrick Marshall[2], Evgeny Mikheev[2], Panpan Zhou[3], Susanne Stemmer[2], and Douglas Natelson[3,4,5]\**

[1]Applied Physics Graduate Program, Smalley-Curl Institute, Rice University, Houston, Texas USA

[2]Materials Department, University of California, Santa Barbara, California USA

[3]Department of Physics and Astronomy, Rice University, Houston, Texas USA

[4]Department of Electrical and Computer Engineering, Rice University, Houston, Texas USA

[5]Department of Materials Science and Nanoengineering, Rice University, Houston, Texas USA

**Abstract**

Heterointerfaces of $SrTiO_3$ with other transition metal oxides make up an intriguing family of systems with a bounty of coexisting and competing physical orders. Some examples, such as $LaAlO_3$/$SrTiO_3$, support a high carrier density electron gas at the interface whose electronic properties are determined by a combination of lattice distortions, spin-orbit coupling, defects, and various regimes of magnetic and charge ordering. Here, we study electronic transport in mesoscale devices made with heterostructures of $SrTiO_3$ sandwiched between layers of $SmTiO_3$, in which the transport properties can be tuned from a regime of Fermi-liquid like resistivity ($\rho \propto T^2$) to a non-Fermi liquid ($\rho \propto T^{5/3}$) by controlling the SrTiO$_3$ thickness. In mesoscale devices at low temperatures, we find unexpected voltage fluctuations that grow in magnitude as $T$ is decreased below 20 K, are suppressed with increasing contact electrode size, and are independent of the drive current and contact spacing distance. Magnetoresistance fluctuations are also observed, which are reminiscent of universal conductance fluctuations but not entirely consistent with their conventional properties. Candidate explanations are considered, and a mechanism is suggested based on mesoscopic temporal fluctuations of the Seebeck coefficient. An improved understanding of charge transport in these model systems, especially their quantum coherent properties, may lead to insights into the nature of transport in strongly correlated materials that deviate from Fermi liquid theory.

Interest in oxide-based heterostructures has intensified in recent years, in large part due to advances in epitaxial film growth and the discovery of a two-dimensional conductive interface between the insulating perovskite materials LaAlO$_3$ (LAO) and SrTiO$_3$ (STO).[1,2] The electron gas at this interface exhibits strong electronic correlations, resulting in competing orders that can be tuned by various external parameters (*e.g.,* temperature, magnetic field, pressure, chemical doping, and electrostatic gating).[3-8] Techniques for growing high-quality films and atomically sharp interfaces of these transition metal oxides (TMOs) and their characterization are central to many ongoing research efforts.[9] The possibility for rich interplay between charge, spin, and orbital ordering makes oxide heterostructure systems excellent tools in the study of correlation phenomena with wide-ranging implications, from understanding high-T$_c$ superconductivity[10] to engineering materials with desirable functional properties.[11-14] Many aspects of these and related oxide heterostructures remain, however, only partially understood, including the nature of quantum coherence in transport, and relationship of coherence with other effects such as spin-orbit coupling, localization (both weak and strong), and charge and magnetic ordering.

Motivating the present study is a body of prior work focused on quantum wells (QWs) in epitaxial STO layers sandwiched between layers of either antiferromagnetic (AFM) SmTiO$_3$ (SmTO) or ferrimagnetic GdTiO$_3$ (GdTO). Those studies, performed using structures with lateral dimensions of ~100 μm$^2$ to 1 cm$^2$, [8,15–17] demonstrated that the transport properties of the SmTO/STO QWs can be tuned from a regime of seemingly Fermi liquid (FL) behavior to a non-Fermi liquid (NFL) regime as the STO thickness is reduced below a critical value of 5 SrO layers, as inferred from the temperature dependence of the resistance. Fitting with the functional form $R = R_0 + AT^n$, a fitting exponent $n < 2$ is interpreted as NFL behavior (in this case, the present NFL samples exhibit $n \sim 5/3$ over a fairly broad temperature range). A subsequent work[18] focused on tunneling spectroscopy in these QW samples, uncovering spectroscopic evidence of a pseudogap at low temperatures in AFM-confined QWs, as well as evidence of coherent transport in these structures.

Here we expand upon those measurements to examine transport in structures of mesoscopic size. We find unexpected, time-dependent voltage fluctuations that emerge at low temperatures for devices of small contact electrode size (~ 250 nm width). These fluctuations were detected between voltage probes in measurements of temperature-dependent resistivity and as a correction to the field-perpendicular-to-plane magnetoresistance (MR). Remarkably, the voltage fluctuations are independent of measurement current, ruling out conventional temporal universal conductance fluctuations (UCF) as a mechanism, and are systematically suppressed with increasing contact spatial size and increasing temperature. The suppression at larger temperature and spatial scales implies that the fluctuations are mesoscopic in nature. We propose a possible origin of the observed behavior in terms of a time-varying Seebeck coefficient within this system.

**Results**

The SmTO/STO/SmTO heterostructures used in these studies were grown by hybrid molecular beam epitaxy (MBE) on LSAT substrates as described elsewhere [8,16,19] (also see the Methods section for additional sample preparation details). Based on prior transport results[17], the STO layer thickness (given in number of SrO layers) was chosen to yield one nominally NFL sample (4 SrO layers) and one nominally FL sample (10 SrO layers) for comparison. The as-grown QW films were patterned into micron-scale test structures with several Ti/Au metal contacts, in two geometries: a Hall bar with lateral dimensions of ≳10 μm, and a long mesa (~3.5 μm wide and ~200 μm long) of QW material with multiple narrow contacts (either nominally all ~ 250 nm width, or in a range of widths from ~0.2 – 5 μm) along its length at variable spacing intervals (~0.1 – 3 μm contact separations), as well as large current contacts located at the mesa ends in the long direction. Top view images of representative devices of each type are shown in Fig. 1, along with a cross-section diagram of the QW stack. These structures were used to measure temperature-dependent resistance, MR isotherms, and time-dependent isotherms to assess low-temperature fluctuations initially observed in the resistance, but later found to be pure voltage fluctuations independent of measurement current.

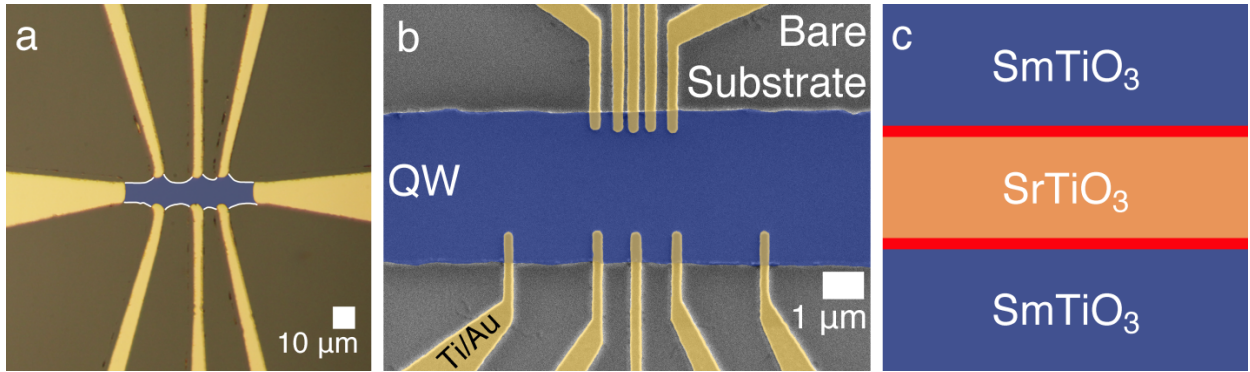

**Fig. 1: Sample device layout.** Top-view images of (a) a Hall bar structure (optical image, QW area highlighted in blue) and (b) a narrow mesa-style sample with variable contact spacing (false-color SEM image; large current lead contacts, not shown, are located beyond the field of view at the left and right ends of the mesa). In both cases, the current flows along the horizontal direction through a mesa of QW structure isolated by etching. Voltage probes located along the QW mesa's edges are used to sense the potential drop. (c) Cross-sectional schematic view (not to scale) showing the structure of a SmTO/STO/SmTO layer stack. Red bars schematically represent the 2d electron gases that form at the SmTO/STO interfaces (though a fraction of the electrons in the $d_{xz}$ and $d_{yz}$ orbitals are delocalized throughout the depth of the well, and overlap of the gases can take place in thin wells)[20].

The temperature dependent resistance (shown in Fig. 2 (a-d)) has a positive slope at high temperatures for both FL and NFL samples, consistent with metallicity, followed by a small low-

T upturn that appears at around 20 K upon cooling. Prior work[16] on similar quantum well structures found that the low-T upturn in resistivity is consistent with interface scattering of charge carriers due to spin fluctuations of moments within the antiferromagnetic insulating SmTO. The sheet resistance at 2 K is ~ 135 Ω/square (FL sample) or ~ 345 Ω/square (NFL sample). At 2 K, the FL sample has an apparent carrier density (estimated from Hall measurements) of ~ $3.2 \times 10^{14}$ /cm$^2$ and a mobility of ~ 140 cm$^2$/V s; by comparison, at 2 K the NFL sample has a carrier density of ~ $9.4 \times 10^{14}$/cm$^2$ and an inferred mobility of ~ 19 cm$^2$/V s (see Fig. S1)). Prior work[17] has noted that the apparent carrier density values inferred from Hall measurements (and subsequently inferred carrier mobility) in these structures are not strictly accurate at low temperatures due to the phenomenon of lifetime separation, which entails different scattering rates governing longitudinal transport and Hall resistances. Thus, it is more accurate to label these measurements as $(eR_H)^{-1}$ rather than carrier density.

For both FL and NFL samples, the resistance is measured in a four-terminal configuration with two large current leads and two small voltage probe electrodes (~ 250 nm wide, as in Fig. 1b), with a constant ac current in the range 10 – 200 nA, sourced using a large series resistance (much larger than the two-terminal device resistance). Unexpected, greatly enhanced fluctuations of the inferred resistance were observed at low temperatures (below ~ 20 K). They are clearly visible in the temperature dependent resistance plots (see Fig. 2 (a-d)), with the fluctuations growing as the temperature is decreased below ~ 20 K, despite the fact that the (average) magnitude of resistivity changes much less in this temperature range. Such fluctuations were not observed in the larger Hall bar type devices (of the style shown in Fig. 1a) to within the background noise level of the measurement, although in the Hall bar devices the overall shape of the R(T) curves is in otherwise good agreement with what is measured using the small-contact samples.

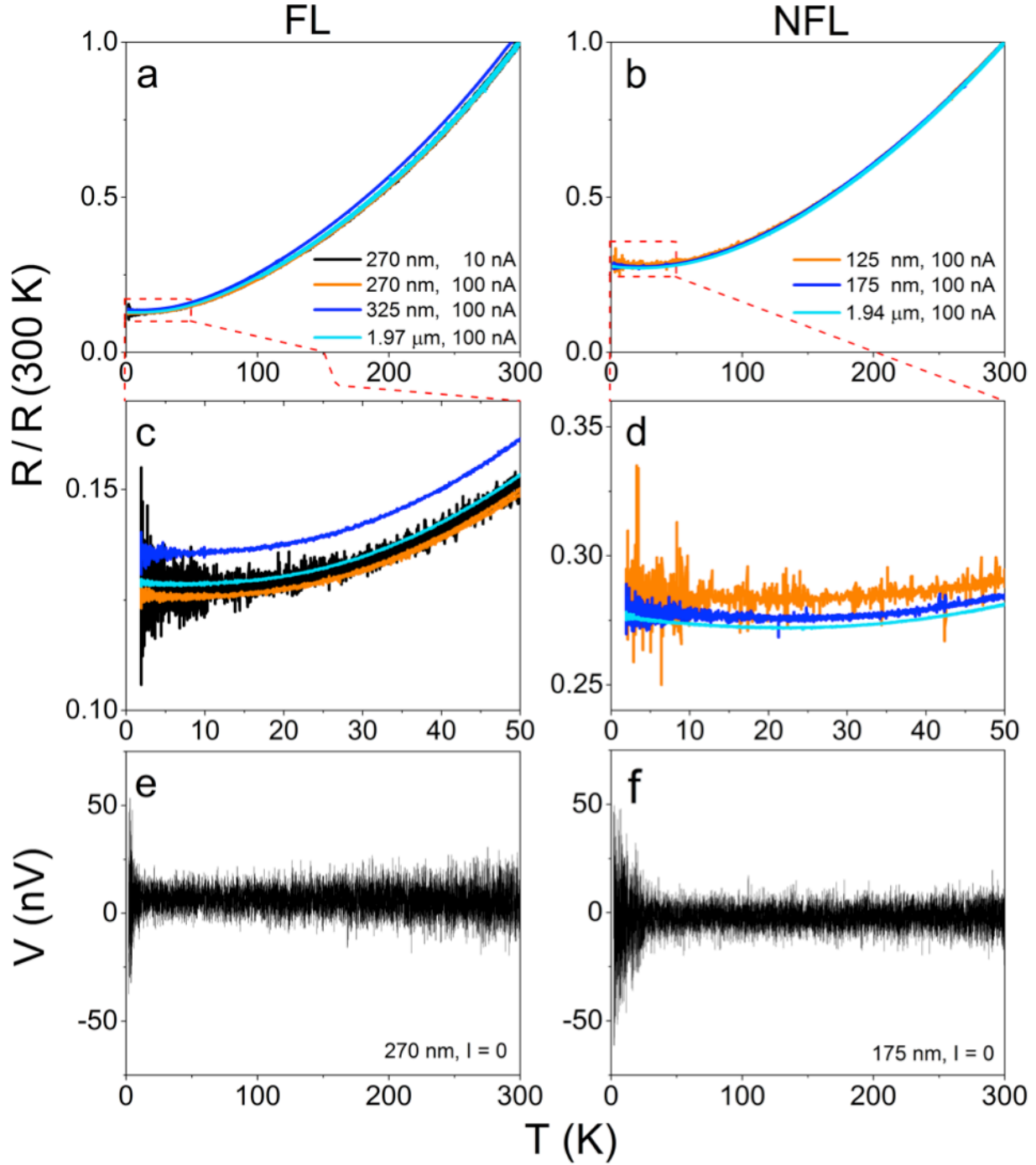

**Fig. 2: Temperature dependence of resistance and voltage fluctuations without current.** Resistance (normalized to T = 300 K values) as a function of temperature for mesa-style samples of both FL (a,c) and NFL (b,d) type, with various separation distances between narrow voltage probe contacts and using various drive currents. Panels (c) and (d) show enlarged views of the region below 50 K, where enhanced fluctuations appear as the temperature is decreased below ~ 20 K. Even when the drive current is zero

(panels (e) and (f)), the measured potential between two narrow voltage probe electrodes (separated by 270 nm for FL or 175 nm for NFL) displays fluctuations that grow with decreasing temperatures, reaching several tens of nV in amplitude at T = 2 K.

This phenomenology is superficially reminiscent of time-dependent universal conductance fluctuations (TDUCF). [21–28] In TDUCF, fluctuating disorder alters the relative phases of coherently interfering electronic trajectories, leading to a time-varying mesoscopic correction to the resistance that is suppressed by ensemble averaging as temperature or length scales are increased. The TDUCF typically have a $1/f$ frequency dependence of spectral power density, and since they are true resistance fluctuations, TDUCF voltage noise power scales with the square of the measurement current.

Several comparison measurements were made using small contacts separated by various distances, and with drive currents ranging from 10-200 nA. In contrast to the expectations for TDUCF, the *voltage* fluctuation magnitude is approximately the same for all measurements, while the inferred *resistance* fluctuation level appears to decrease with larger drive current or larger contact separation distance due to the larger measured voltage drop (*i.e.*, the ratio of the voltage fluctuation size to the total voltage drop, δV/V, decreases). In contrast to the fluctuating potential, the simultaneously measured drive current and its fluctuation level are found to be stable under all measurement circumstances, with no appreciable change at low temperatures.

We further confirm the lack of dependence of the fluctuations on the drive current by measuring the voltage drop between two electrodes as a function of temperature, *in the absence of any drive current* (the current terminals are left electrically floated), as shown in Fig. 2(e,f). Just as for the resistance measurement, here the potential fluctuations increase dramatically at low temperatures for both FL and NFL samples, reaching several tens of nV.

We examined the voltage fluctuations in further detail using frequency domain measurements.[29,30] We measured the amplitude of the potential difference between various pairs of electrodes, under zero applied current, as a function of frequency using a fast Fourier transform (FFT) spectrum analyzer, over a range of 0.5 - 25 Hz. For each measurement, 500 individual spectra were collected and averaged to reduce background noise contributions. As shown in the log-log plots of Fig. 3, the measured spectra have decreasing amplitude contributions at higher frequencies for both FL and NFL samples, following a rough 1/f relationship for most traces. For small size contacts (~250 nm wide voltage probes), the measured amplitude was largest at low temperatures, decreasing by more than an order of magnitude at room temperature. By comparison, for large contacts (the current leads at the ends of the mesa, ~ 3.5 μm wide), the fluctuation spectrum at 2 K was in reasonable agreement in magnitude with the 300 K data for small contacts, approximating the noise floor of our measurement system (see Fig. S5). These trends indicate that the fluctuations are most prominent in contacts of small size, and the separation distance between the small contacts seems to have little effect on the voltage fluctuation amplitude at low temperatures, as shown in panels (b) and (c) of Fig. 3 for two pairs of narrow contacts on an NFL sample, separated by > 10× different distances.

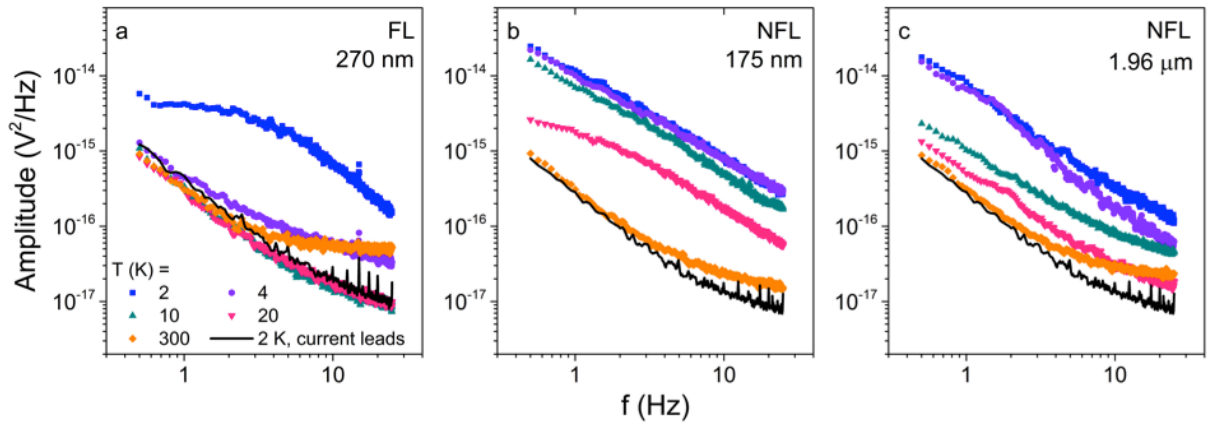

**Fig. 3: FFT spectra of fluctuation amplitude at low frequencies and various temperatures.** Log-log plots of fluctuation amplitude (at zero drive current) as a function of frequency, collected using *narrow* ~250 nm wide voltage probe contacts (solid points) on (a) FL sample with contact separation ~ 270 nm, (b) NFL sample with contact separation ~ 175 nm, and (c) NFL sample with contact separation ~ 1.96 μm. For comparison, each plot also includes the spectrum collected using the large contacts at the two ends of the mesa at T = 2 K (black solid line). This trace is of the same magnitude as the room-temperature curves collected using small voltage probe contacts. Note that the spectral traces in solid points in (b) and (c) are similar in magnitude despite the > 10× difference in contact separation distance, which supports the idea that contact size is the most important factor in determining the fluctuation amplitude.

To further constrain possible mechanisms behind these observations, we also consider the transport properties under application of an external magnetic field. MR curves were measured at selected temperatures for both NFL and FL samples, using both small and large voltage probe contacts separated by various distances in the range of ~ 100 nm – 10 μm (see supporting information for in-depth temperature-dependent results). For small voltage measurement contacts, and most prominently in the NFL case (also for FL samples, but to a lesser extent), the MR shape is overlaid with large time-varying fluctuations (due to the potential fluctuations), causing a mismatch between the portions of the curve sweeping up and down in field. Even at relatively large currents (200 nA), under which conditions the temporal voltage fluctuations feature less prominently in the resistance (due to the larger average voltage drop compared to the roughly fixed fluctuation size at a given temperature), Fig. 4 shows that sweeping the field back and forth several times between ± 9T produces a collection of MR curves that presumably should be identical, but differ with each other in their detailed structure (also see additional data in Figs. S3 and S4). We further note that there is little effect of an applied magnetic field of up to 9 Tesla on the *magnitude* of the time-dependent potential fluctuations for either FL or NFL samples. This, too, is inconsistent with the Feng-Lee-Stone TDUCF theory prediction of a relative reduction in noise by a factor of two when a sufficiently strong magnetic field is applied to a disordered conductor and time reversal symmetry is broken (where the required field strength is proportional to the inverse square of the phase coherence length $L_\phi$).[25–28,31] These two behaviors are not in agreement with the typical observations of universal conductance fluctuations, and suggest the need for a better understanding of transport in these QW systems.

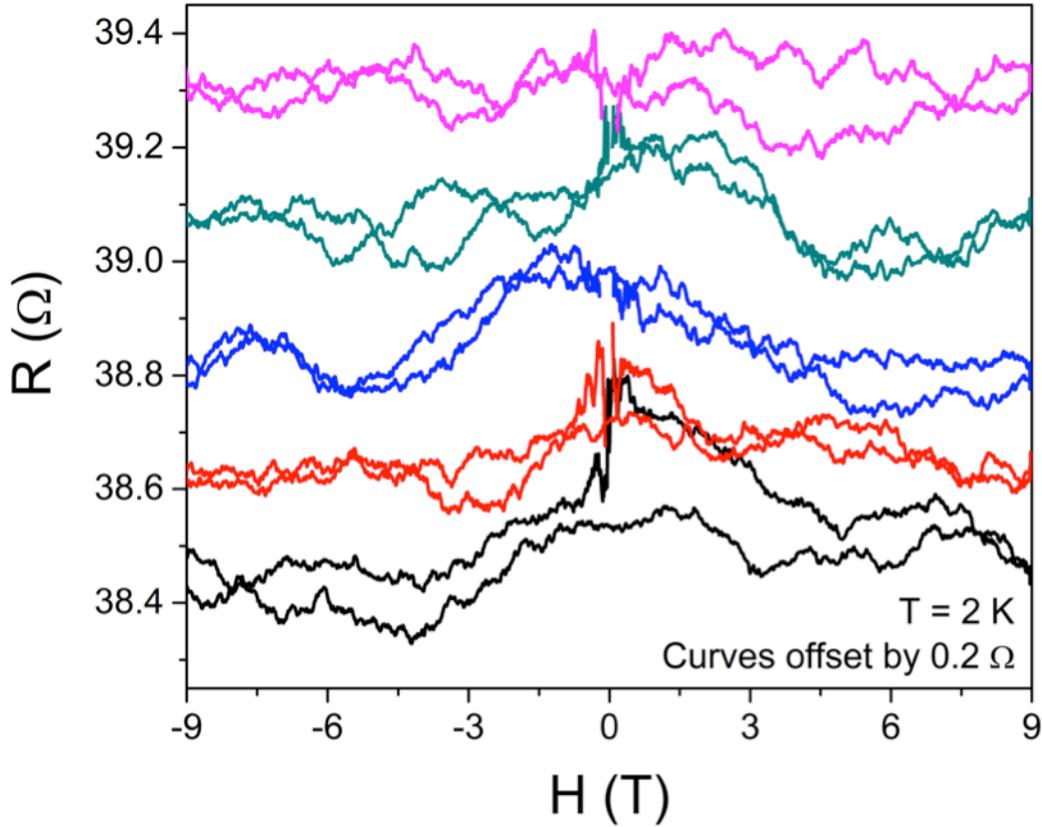

**Fig. 4: Magnetofingerprint-like MR traces at 2 K**. Five repeated four-terminal MR sweeps for the NFL mesa sample with voltage probes separated by ~125 nm, taken while the temperature was held stable at 2 K and with the current fixed at I = 200 nA. Curves are offset by 0.2 Ω for clarity, starting from lowest (black) trace. For some curve sections, the fluctuations retrace closely upon sweeping from zero to high field and back to zero, which is a characteristic feature of a "magnetofingerprint." However, the effects of time-dependent fluctuations in the potential landscape result in successively acquired MR curves that do not match one another in their detailed fluctuations.

To verify that the large potential fluctuations are indeed due to the intrinsic QW sample properties, it is necessary to rule out other possible origins of fluctuations, which could include instrumental noise, thermal noise, contact effects, and conventional UCF. We first consider the contribution of voltage preamplifier noise (including current noise on the preamplifier voltage inputs), the magnitude of which depends on the impedance of the measured sample. The preamplifier is a Stanford Research Systems SR560, for which in low-noise mode, the input noise floor is about 10 nV/√Hz for R < 1kΩ at 10 Hz (see Fig. S5). To rule this out as a significant source of noise, we collected time dependent fluctuation data on two kinds of control samples, mounted on the PPMS sample holder and cooled to the same measurement temperatures as for the QW samples. These control samples consisted of either commercial metal film resistors with nominal values of 100 Ω, 1 kΩ, and 10 kΩ (whose resistances were chosen to simulate the two- or four-terminal QW device resistances, and were nearly temperature-independent over the range 2 – 300 K), or an evaporated strip of $Au_{0.6}Pd_{0.4}$ metal film (fabricated on an oxidized silicon wafer by e-beam lithography, e-beam evaporation, and liftoff; dimensions about 10 μm x 200 μm x 50 nm) with a room temperature resistance of ~ 88 Ω, falling to ~ 78 Ω at 2 K. The measurement apparatus and control settings (*e.g.*, amplifier gain) were the same as those used for QW studies. These tests showed no dependence of voltage fluctuation level on control sample temperature. For completeness, we also note that the temperature as measured by the internal cryostat thermometer is stable to typically < 10 mK during isothermal measurements.

We also consider the possible role of contact resistance and current noise on the preamplifier inputs in the measured voltage fluctuations enhancement at low temperatures, but find that this can reasonably be ruled it out as the primary source of the large fluctuations. Measurements of the 2-terminal resistance as a function of temperature (Fig. S6) show that the sum of contact and channel resistances does not change greatly over the temperature range in which the potential fluctuations are observed to change by factors of several. For the FL sample, the 2-terminal resistance at 2 K is a factor of ~4 smaller than at room temperature, and the increase in 2-probe resistance from 10 K to 2 K is only ~ 10%, while the potential fluctuations grow by more than an order of magnitude over that temperature range (Fig. 3a). Given the relatively modest change in 2-terminal resistance over the temperature range studied in both FL and NFL cases, we posit that contact resistance cannot reasonably account for the dramatic increase in voltage fluctuations at low temperatures.

Similarly, the expected Johnson-Nyquist (J-N) noise of the QW sample must also be considered, accounting for both decreasing temperature and increasing two-terminal contact resistance at low temperatures. Using the value of the measured sample resistance (a typical value for four-terminal resistance would be ~100 Ω at low temperature, or in the range of a few kΩ for two-terminal resistance, which takes contact effects into account) and the frequency bandwidth $\Delta f$ of the measurement, the expected J-N noise $V_n = \sqrt{4k_B T R \, \Delta f}$ can be calculated. At a temperature of 2 K, with a frequency bandwidth of 25 Hz for the FFT measurements, a 100 Ω resistor is expected to have $V_n \sim 0.5$ nV. This is significantly smaller than the actual measured magnitude of voltage fluctuation of several tens of nV. Even a growing 2-terminal contact resistance at low temperatures cannot explain the observed fluctuations as $T$ is decreased in terms of Johnson-Nyquist noise.

**Discussion**

Taking into account the data presented above, we now seek a self-consistent explanation for the two groups of phenomena observed in the QW samples: the large, low-temperature, current-independent mesoscopic voltage fluctuations observed as a function of time, as well as the magnetofingerprint style resistance (or conductance) fluctuations measured as a function of magnetic field. Recent findings[18] imply that below ~ 20 K a pseudogap forms in the NFL samples and a zero-bias anomaly appears for the FL samples (the latter consistent with weak depletion of the density of states near the Fermi level), and it makes sense to consider whether these phenomena are relevant to our observations. There is no evidence that the fluctuations are due to proximity to a regime of strong localization, given the samples' low sheet resistance, positive $R(T)$ slope for the FL sample down to 2 K, and very slight low-$T$ upturn in resistivity for the otherwise metallic NFL sample.

The presence of voltage fluctuations even in the absence of an applied current strongly constrains the possibilities. Since both types of samples measured here exhibit similar magnitudes of potential fluctuation, we conjecture that the fluctuations are likely a feature of the conductive SmTO/STO interface and its etching-induced defects, rather than a consequence of FL-to-NFL crossover with STO layer thickness. An explanation involving transduction of a fluctuating thermopower is most consistent with the observations. The difference in Seebeck coefficients of dissimilar metals can generate a net voltage in the absence of any bias current. If some microscopic mechanism causes local temporal fluctuations in the Seebeck coefficient $S$ of the QW electron gas, this would be measured as a fluctuating voltage by the sensing electrodes. These SmTO/STO/SmTO QWs are known to display a relatively large thermopower, on the order of 1-10 μV/K at temperatures below 50 K,[20,32] and mesoscopic fluctuations in the local chemical potential could credibly manifest as a fluctuating thermovoltage on the order of tens of nV (corresponding to a fluctuation of $S$ on the order of 1%). While full, quantitatively calibrated measurements of the thermopower of the QW/voltage contacts are beyond the scope of this work, preliminary measurements using optical heating (see supporting information text and Fig. S8) show that such contacts can readily exhibit large thermoelectric voltages. Prior work has established that in mesoscopic devices, fluctuation phenomena analogous to what is often observed in electronic transport also appear in thermoelectric transport. A previous experimental study presented results of magnetic-field dependent resistance and thermopower as direct evidence that the Onsager reciprocity relations extend to the case of thermoelectric transport,[33] resulting in mesoscopic fluctuations in thermoelectric coefficients, and there is corresponding theoretical support as well.[34,35] An additional control experiment was performed using analogous QW structures that were prepared either with or without etching to define a mesa (see Supporting Information section VII and Fig. S9). The structures fabricated without etching show no evidence of low-temperature temporal fluctuations, whereas the etched structures show similar behavior as observed in the SmTO/STO/SmTO samples described above.

A fluctuating Seebeck coefficient with a microscopic origin of fluctuating chemical potential due to two-level defects[36–38] in the etched nanostructures is reasonable in the context of an emerging pseudogap at low $T$. The pseudogap in these QW structures emerges over the same temperature range as the increase in voltage fluctuations, as $T$ is reduced below 20 K.[18] The transfer of states away from the Fermi level during pseudogap formation and the resulting energy dependence of the density of states is both likely to enhance potential fluctuations due to defects and to enhance the Seebeck coefficient. An analogous response has been observed previously in germanium single crystals, where carrier density fluctuations result in a 1/$f$ noise spectrum of the Seebeck coefficient, even without a dc current.[39] In the case of our QW samples, such carrier density fluctuations could be imagined to occur on one or both sides of the heterointerface, though within the metallic QW system with its carrier density of ~ $10^{14}$/cm$^2$, screening should take place on very short length scales. Regarding deviations from pure 1/$f$ frequency dependence, while a *1/f* functional form of distribution of fluctuators is very common, this is not *a priori* required; if there were only a single type of fluctuating defect introduced by etching, one would not expect a broad frequency distribution at all.

Local fluctuations in Seebeck coefficient would be consistent with suppression of the voltage fluctuations in larger contacts, as increasing contact size would lead to ensemble averaging of local fluctuations. In addition to direct measurements of thermopower, another possible test of this picture would be low-temperature Kelvin probe force microscopy, an approach that can locally image spatial and temporal variations in the chemical potential. This specialized technique is, however, beyond the scope of the present work.

The magnetofingerprint MR traces in Fig. 4 and Fig. S3, S4 further support the idea that mesoscopic fluctuations are detectable in these oxide QW systems, and that the potential landscape is varying in time. In the presence of static disorder, a retraceable "magnetofingerprint" is the signature of UCF as a function of external magnetic field (MFUCF). Long established as an indicator of coherence in transport on the mesoscopic scale, [21–24] MFUCF have been reported in other oxide structures.[40–43] As shown in the figures, the magnetofingerprint is fairly retraceable on the timescale of tens of minutes, but time-varying disorder (due to the presumed two-level fluctuators responsible for the voltage fluctuations) scrambles the MR on the time scale of hours.

**Conclusions**

We have studied the transport properties of interface quantum wells in SmTO/STO/SmTO heterostructures using mesoscopic devices with micro/nanoscale electrodes. Unexpected time-dependent voltage fluctuations are observed in devices with the smallest voltage contacts at low temperatures. These fluctuations are found to be suppressed with increasing temperature and contact size, and are independent of the drive current. The suggested mechanism most consistent with the data involves a fluctuating Seebeck coefficient that is transduced as a fluctuating potential at low temperatures. The presence of mesoscopic fluctuations and time-varying disorder is further supported by the observation of "magnetofingerprint" magnetoresistance at the lowest temperatures that is short-time retraceable but shows strong variation on the timescale of hours. This work highlights the need for an improved understanding of the mesoscopic electronic properties in complex oxides that are not necessarily conventional Fermi liquids.

**Methods**

SmTO/STO/SmTO quantum well samples were grown by hybrid MBE on commercial LSAT substrates, as reported elsewhere.[19,44,45] The STO layer thickness, described in number of SrO layers, was verified using cross-sectional scanning transmission electron microscopy (STEM).

Devices for transport measurements (Hall bars and mesa samples) were fabricated from the as-grown QWs by a combination of e-beam and photolithography to define contacts for metallization (e-beam evaporation of 5 nm Ti and 50 nm Au), followed by room-temperature directional reactive ion etching (RIE) with chlorine to isolate well-defined regions of QW sample for study. For the etching process, the metal contacts were protected with either 35 nm of $Al_2O_3$ deposited by e-beam evaporation (for samples with several small contacts at variable spacing) or Shipley S1813 photoresist (for larger Hall bar samples).

Transport measurements were performed using low-frequency lock-in techniques in a Quantum Design Physical Property Measurement System (PPMS). After processing the as-grown samples into test structures, temperature-dependent resistance and variable-temperature Hall measurements were performed and their results compared to those obtained in advance with the unprocessed films (which were measured in Van der Pauw geometry), to ensure that the QW properties had not changed over time during the device fabrication. Since the QW samples are known to degrade when baked at high temperatures in ambient atmosphere, all processing was limited to temperatures below 120°C, and no evidence of degradation was observed when comparing transport data taken before and after processing.

**Associated Content**
**Supporting Information**

The Supporting Information is available free of charge on the ACS Publications website at DOI:[insert DOI here].

Transport characterization of the as-grown quantum well samples, additional magnetoresistance data, determination of measurement system noise floor, 2-terminal resistance vs. temperature measurements, potential fluctuation level dependence on contact size, preliminary thermophotovoltage measurements, and control experiment results using $GdTiO_3/SrTiO_3/GdTiO_3$ structures prepared without etching.

**Author Information**

**Corresponding Author**
*E-mail: natelson@rice.edu
**Notes**
The authors declare no competing financial interest.

**Acknowledgements**
W.J.H., P.Z., and D.N. gratefully acknowledge support from the US DOE Office of Science/Basic Energy Sciences award DE-FG02-06ER46337, and thank P. Zolotavin for help with the photothermovoltaic measurements. Work at UCSB was supported by the U.S. Army Research Office (Grant No. W911NF-14-1-0379).

Supporting Information for:

# Potential Fluctuations at Low Temperatures in Mesoscopic-Scale SmTiO$_3$/SrTiO$_3$/SmTiO$_3$ Quantum Well Structures

*Will J. Hardy[1], Brandon Isaac[2], Patrick Marshall[2], Evgeny Mikheev[2], Panpan Zhou[3], Susanne Stemmer[2], and Douglas Natelson[3,4,5]\**

[1]Applied Physics Graduate Program, Smalley-Curl Institute, Rice University, Houston, Texas USA

[2]Materials Department, University of California, Santa Barbara, California USA

[3]Department of Physics and Astronomy, Rice University, Houston, Texas USA

[4]Department of Electrical and Computer Engineering, Rice University, Houston, Texas USA

[5]Department of Materials Science and Nanoengineering, Rice University, Houston, Texas USA

\*email: natelson@rice.edu

# I. Characterization of as-grown QW samples

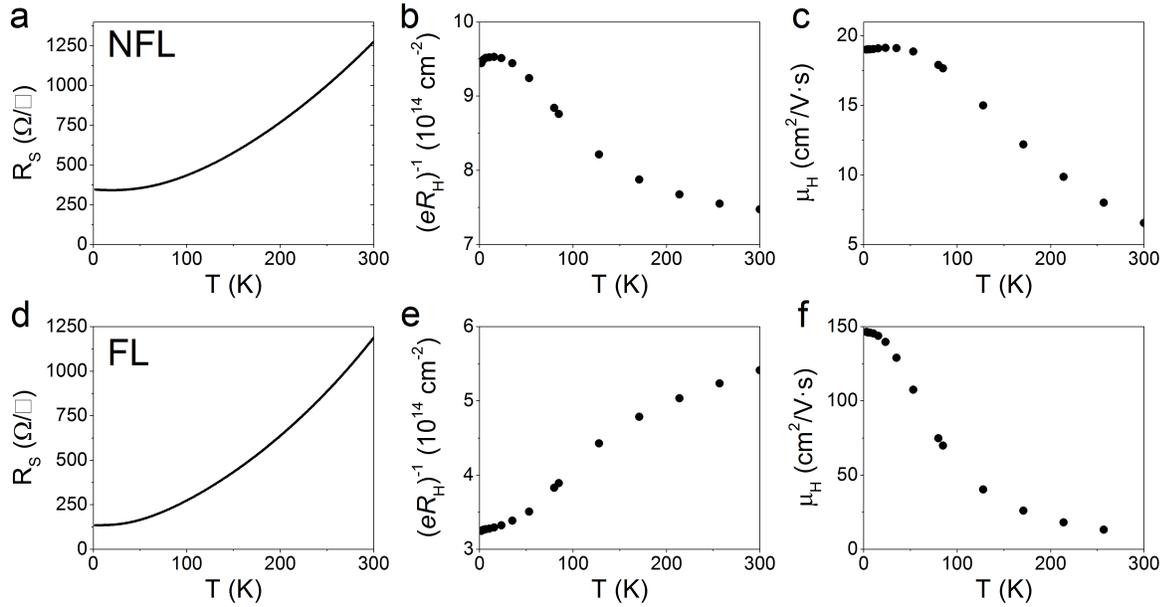

**Fig. S1**: Initial characterization of the as-grown NFL sample (top row, a-c) and FL sample (bottom row, d-f) (each of lateral dimensions 1 cm$^2$) in van der Pauw configuration showing temperature dependence of sheet resistance (a,d), apparent carrier density $(eR_H)^{-1}$ (b,e) and mobility (c,f) inferred from Hall and resistivity measurements. As discussed in the main text, the apparent carrier density (and inferred mobility) values obtained from Hall measurements are not accurate at low temperatures due to the phenomenon of lifetime separation.

# II. Additional magnetoresistance data

The NFL samples (Fig. S2, panels a-d) show a negative MR at low temperatures with a cusp at zero field. This curve shape flattens noticeably as the temperature is increased to 75 K, and becomes weakly positive by 100 K. The MR magnitude is less than 1% at 2 K and less than 0.1% at 100 K. In contrast, for the FL samples (Fig. S2, panels e-g), the MR is positive at all temperatures and is nearly linear at 100 K, with magnitude ~ 0.25%, increasing to ~ 0.5% by 2 K. As the temperature is decreased below ~ 100 K, an approximately flat region appears at low magnetic fields that widens with decreasing temperature, with an upturn at a field scale of ~ 3.7 T at T = 2 K (though the precise field value of the upturn is difficult to determine due to the fluctuations).

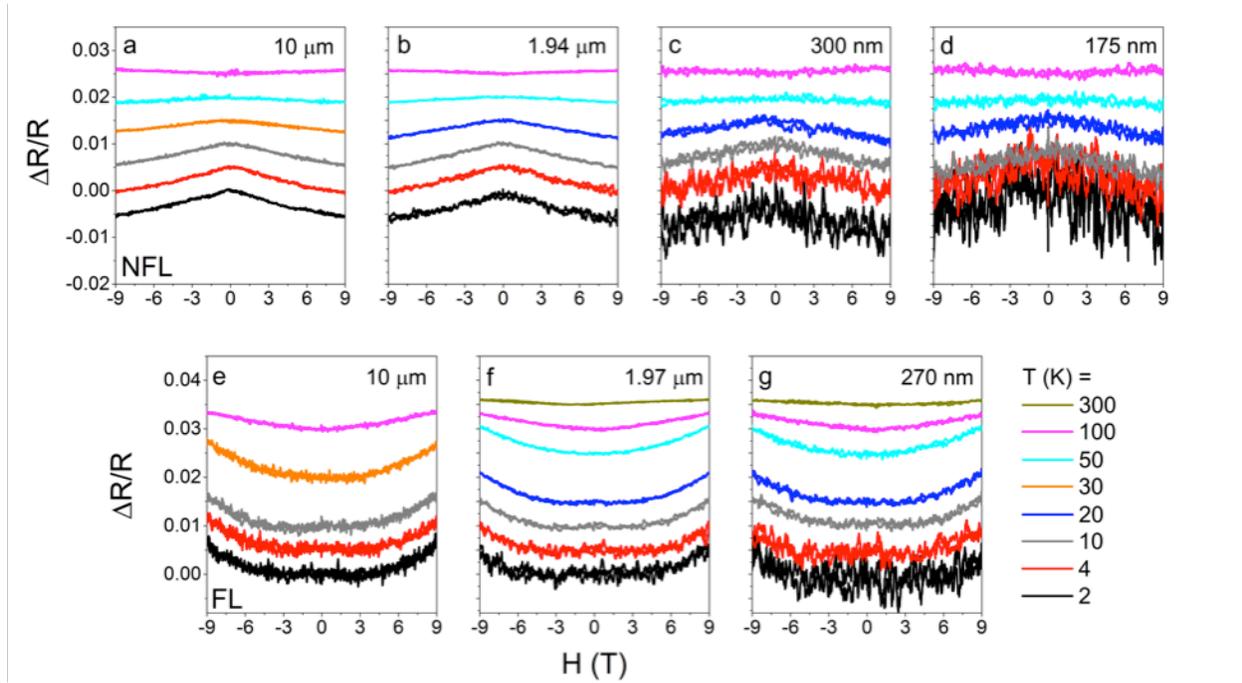

**Fig. S2: Magnetoresistance isotherms.** Curves are offset by 0.005 per temperature value for clarity. Top row (a-d): MR curves at selected temperatures for NFL samples of various contact separation distances (I = 100 nA). The MR is negative at low temperatures and becomes slightly positive by 100 K. Bottom row (e-g): MR curves at selected temperatures for FL samples of various contact separations (I = 100 nA). The MR is positive at all measured temperatures, with an approximately flat region at low field below 100 K, widening as the temperature is decreased. For both NFL and FL samples, the apparent fluctuation level is nearly constant at different temperatures for curves measured using the relatively large ~ 5 μm wide voltage contacts of a Hall bar device as in (a) and (e), whereas (b-d) show that for various separation distances on a mesa-style sample with ~ 250 nm wide voltage probes, the fluctuation level increases significantly with decreasing temperature at low temperatures. Although the voltage fluctuation level is approximately the same for a given contact size, regardless of the separation distance, that distance determines the total voltage drop and thus, the relative contribution $\delta V/V$ of the fluctuation amplitude to the resistance signal. Curves taken using longer separation distances therefore look less noisy at low T when plotted as four-terminal resistance.

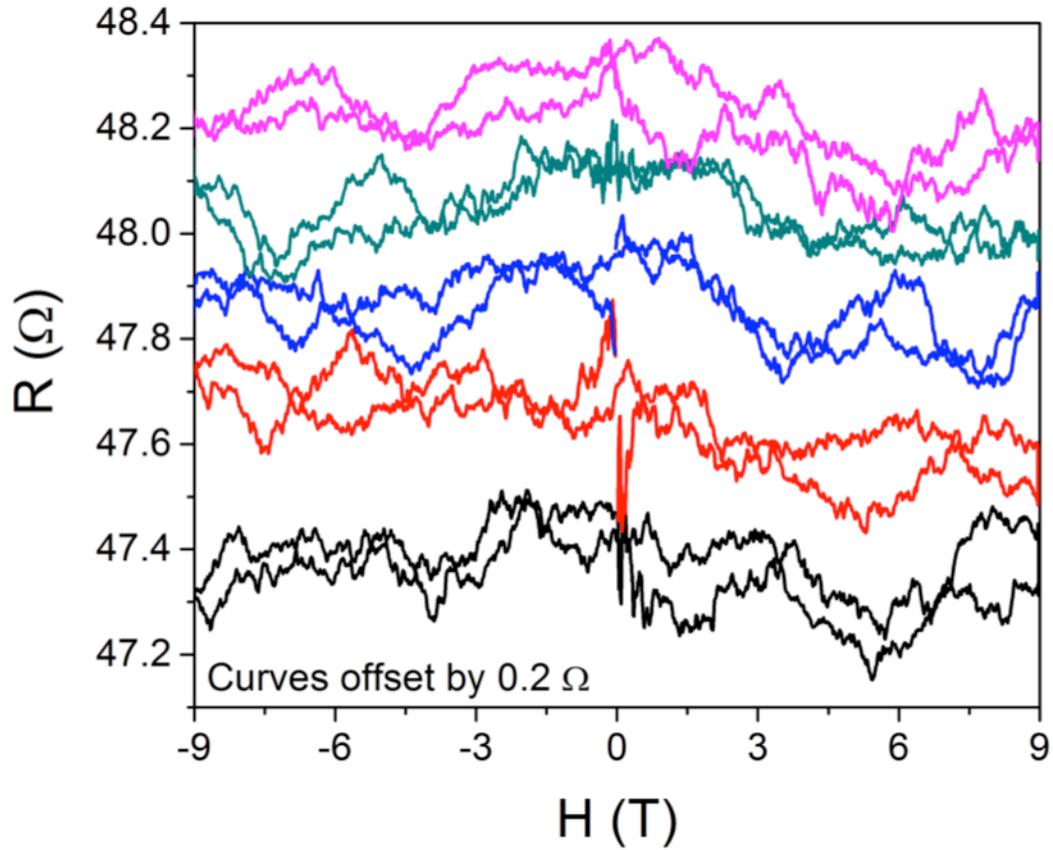

**Fig. S3**: Five repeated MR sweeps for the NFL device, 175 nm contact separation distance. The data were taken consecutively while the temperature was held stable at T = 2 K and the current fixed at I = 100 nA. The effects of time-dependent fluctuations result in MR curves that do not match one another in their detailed fluctuations. Curves are offset by 0.2 Ω for clarity, starting from lowest (black) trace.

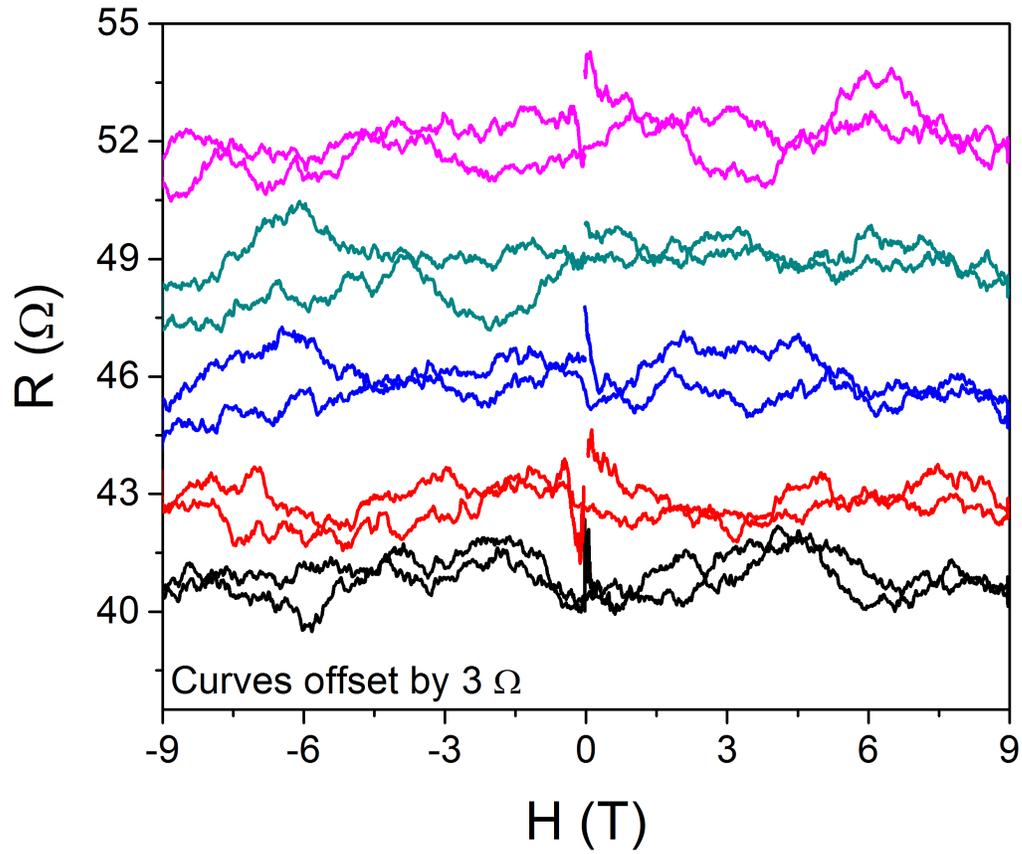

**Fig. S4**: Five repeated MR sweeps for the NFL 125 nm contact separation distance taken while the temperature was held stable at T = 2 K and the current fixed at I = 10 nA. The effects of time-dependent fluctuations result in MR curves that do not match one another in their detailed fluctuations. Curves are offset by 3 Ω for clarity, starting from lowest (black) trace.

## III: Determination of Measurement System Noise Floor

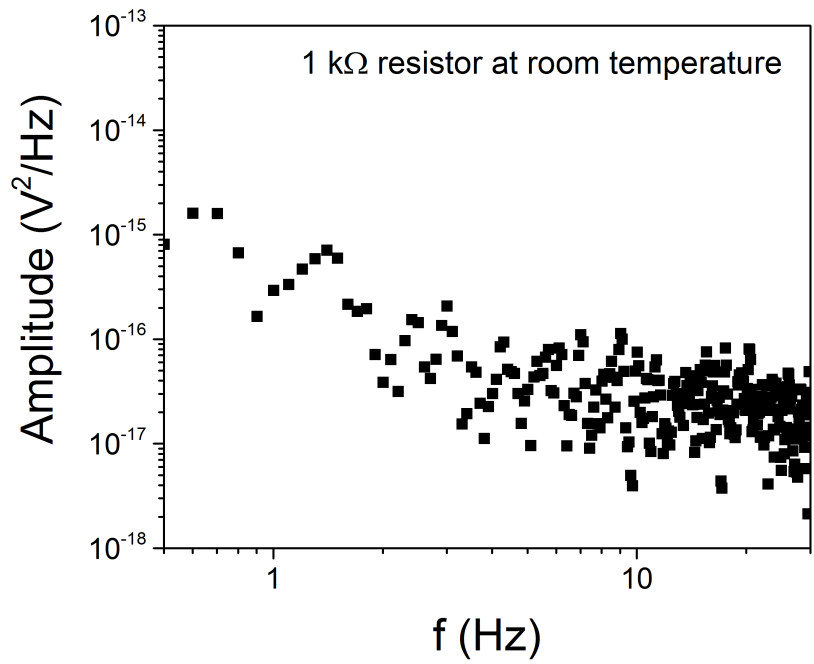

**Fig. S5:** Noise floor of measurement setup, verified using a 1 kΩ resistor mounted in a closed metal box at room temperature. These noise values are consistent with expectations from the technical specifications of the voltage preamplifier.

## IV. 2-terminal resistance vs. temperature measurements

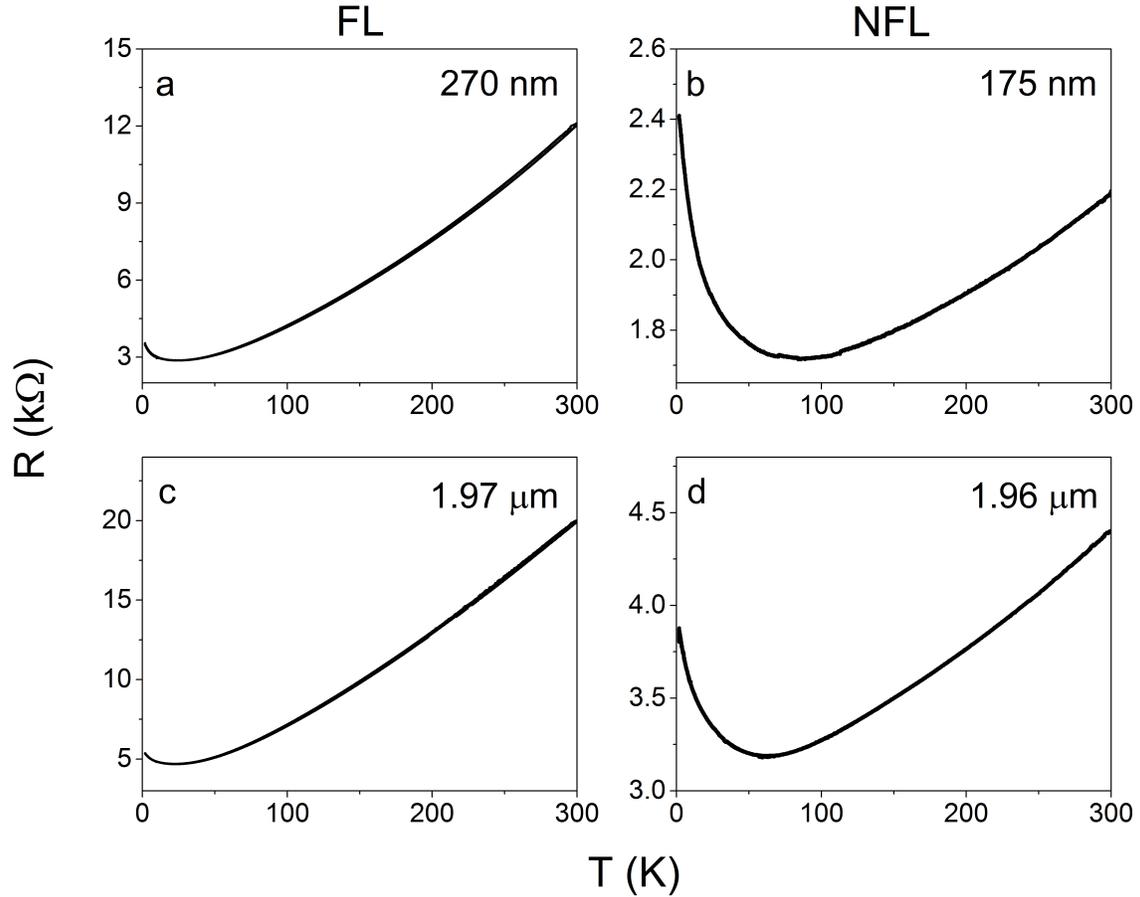

**Fig. S6**: Two-terminal resistance versus temperature plots for FL sample (left column, (a) and (c)) and NFL sample (right column, (b) and (d)), for two different contact separation distances on each mesa-style sample. For the FL sample, the 2-terminal resistance drops by a factor of ~4 over the displayed temperature range, and the increase in 2-probe resistance from 10 K to 2 K is only ~ 10%. For the NFL sample, the 2-probe resistance is not monotonic and increases below ~ 100 K to a value ~ 10% higher or lower than at room temperature, depending on the contact set used. However, neither FL nor NFL sample shows a sudden, extreme change in 2-probe resistance occurring near 10 K, when the large potential fluctuations turn on.

## V. Potential Fluctuation Level Dependence on Contact Size

To further investigate the observation that contacts of smaller size exhibit relatively larger potential fluctuations at low temperatures compared to large contacts, we fabricated a FL mesa-style sample with contacts of various designed widths, in the range ~0.2 – 4.9 μm. We note that accurate determination of contact area is difficult because the *geometric* area measured using, *e.g.*, SEM images may not reflect the *true* contact area over which charge can transfer between the metal contact and the QW. Fig. S6(a) shows the fluctuating voltage measured with various pairs of these contacts as a function of temperature, without any drive current. For pairs of contacts with progressively smaller sums of their widths, enhanced fluctuations appear below ~ 20 K. Fig. S6(b-d) show the roughly linear FFT spectra measured at various temperatures using the same contact pairs as for panel (a). At 2 K, the smallest contacts show a fluctuation amplitude about one order of magnitude larger than for the largest contact set. As the temperature increases to 20 K, all the contact pairs' FFT spectra converge to have more similar amplitudes.

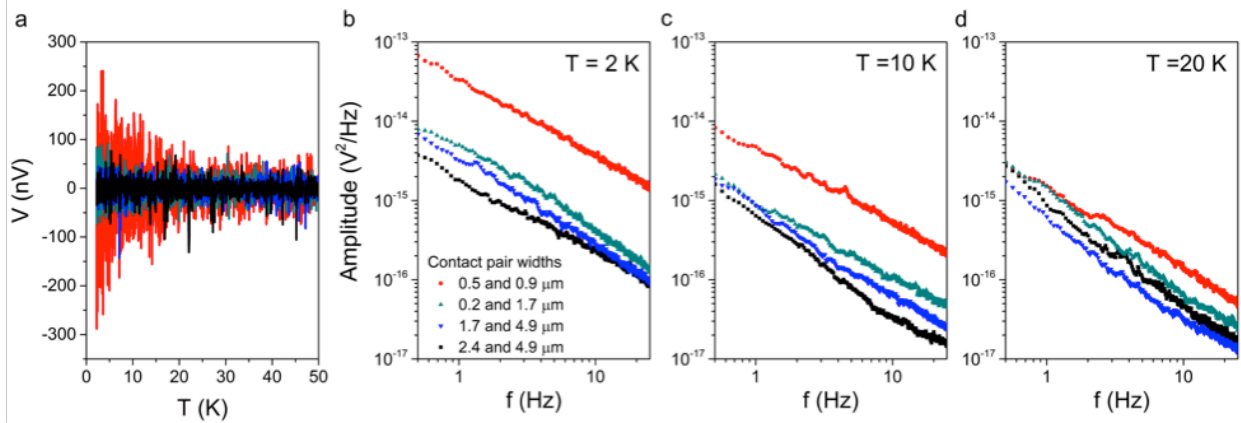

**Fig. S7: Fluctuations measured with different contact sizes.** (a) Potential fluctuation measured as a function of temperature (with zero drive current) collected on a mesa-style FL sample using pairs of contacts of various widths. (b-d) Log-log plots of voltage fluctuation amplitude (at zero drive current) as a function of frequency for the same contact pairs as in (a). At 2K, there is an enhancement of fluctuation

amplitude with decreasing contact size, and as the temperature is increased, the overall fluctuation amplitude decreases for all contact pairs, and the curves for all contact pairs reach similar magnitudes.

## VI. Preliminary Thermovoltage Measurements

Quantitative measurements of the thermopower in mesoscale devices, though technically challenging and beyond the scope of the present work, could be performed to control and measure the local heating, perhaps using lithographically patterned on-chip heaters and thermometers. As a simple alternative demonstration of the importance of thermopower in these structures, preliminary photothermovoltage measurements have been performed in order to check for a resolvable thermally-induced potential difference between two contacts. One of the same mesa-style samples used for transport measurements (NFL type) was mounted in an optical cryostat and cooled to a substrate temperature of 3.5 K. We raster-scanned a 785 nm cw laser focused to a ~ 1 um spot over the sample while measuring the potential between a small 250 nm-wide voltage probe and a grounded large current contact. This local optical heating through absorption, which causes a corresponding induced thermophotovoltage, is most significant when the laser spot is positioned directly over the small voltage probe, with a maximum value of ~ 250 µV when the laser power at the sample is ~ 10 mW (see Fig. S7). Based on measurements of the thermopower in these QWs ($S$ ~ 1-10 µV/K at low temperatures, and likely on the lower end of this range, especially below 10 K),[20,32] the temperature increase due to laser heating can be estimated to be within the range 25-250 K. Conversely, this would imply that a transduced voltage fluctuation of several tens of nV would correspond to a fluctuation of $S$ of approximately <1%. A more well-calibrated method of heating would be required to extract accurate thermopower data, but this method serves as a convenient demonstration of the importance of the Seebeck effect in our QW samples.

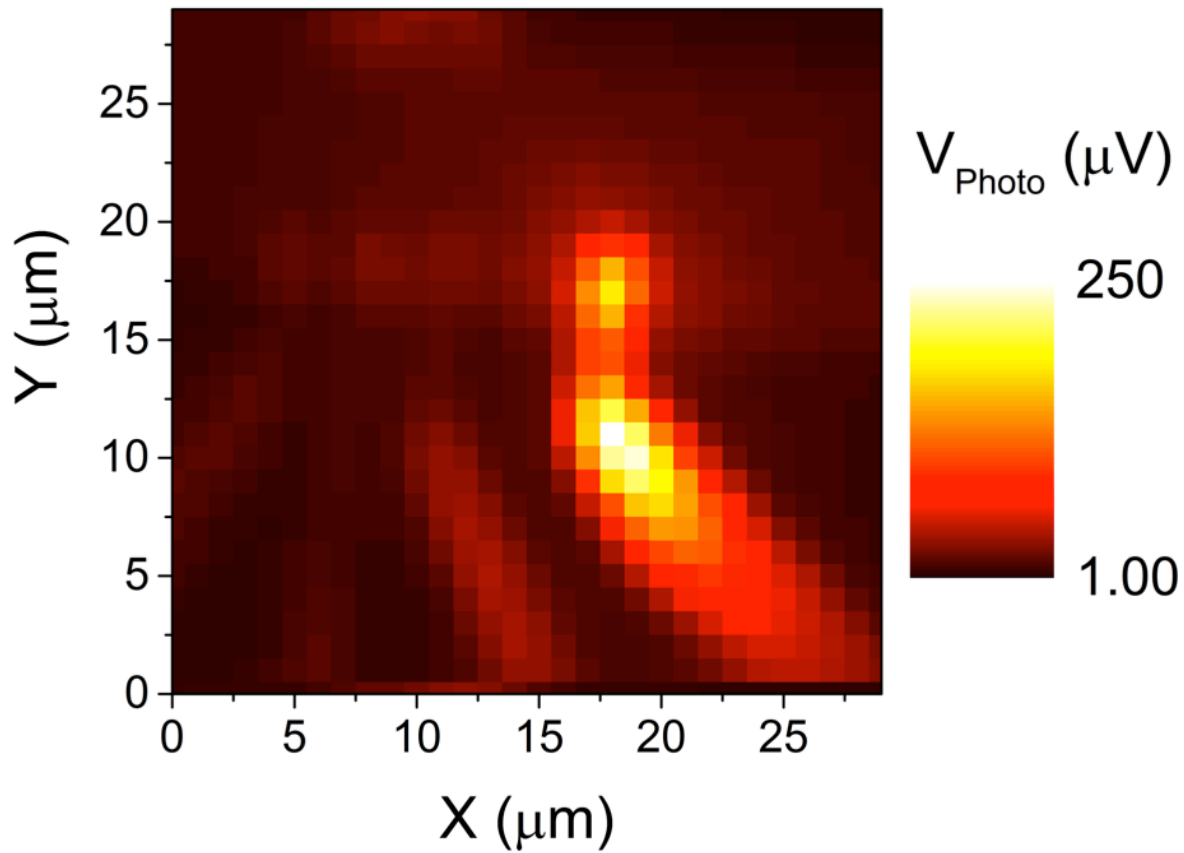

**Fig. S8: Photothermovoltage mapping of FL mesa-style sample.** The map was produced by raster-scanning a 785 nm cw laser (~10 mW at the sample) over the sample at T = 3.5 K while measuring the photovoltage $V_{photo}$ between a narrow voltage contact (brightest area) and a large grounded current lead (not pictured). The contact geometry is the same as shown in Fig. 1(b). The shapes of three other narrow contacts are faintly visible in the bottom left quadrant, as is the horizontal edge of the QW mesa near the middle, but the rightmost voltage probe clearly yields the largest photovoltage when the laser is scanned over it.

## VII. Control Experiment using GdTO/STO/GdTO Structures Prepared Without Etching

We performed an alternate control experiment in a similar system, avoiding the etching step believed to induce defects. We used analogous QW structures made with GdTiO$_3$/SrTiO$_3$/GdTiO$_3$,[8,20] grown with the same layer thicknesses as the SmTO/STO/SmTO samples discussed in the manuscript. Structures were fabricated on pristine QW material by e-beam lithography to form closely-spaced narrow electrodes (inset of Fig. S9a), allowing the current flow to be (primarily) confined to a limited area, but *without* defining a mesa and channel region by etching. In this configuration, no enhancement of time-dependent voltage fluctuations is observed at low temperatures in any measurement, including the temperature-dependent resistivity (Fig. S9a), the temperature-dependent voltage between two narrow contacts measured at zero current (Fig. S9b), and the low-temperature magnetoresistance. A second set of GdTO/STO/GdTO devices was also fabricated in the etched "mesa-style" configuration (the same as described in the main text), and in that case, low-temperature fluctuations are indeed observed, and are very similar to what we discuss in the case of SmTO/STO/SmTO. These results support the idea that etching-related defects are the real culprits of the fluctuation behavior, and that the large Seebeck coefficient serves as a likely mechanism to amplify the effects of local fluctuations in the chemical potential.

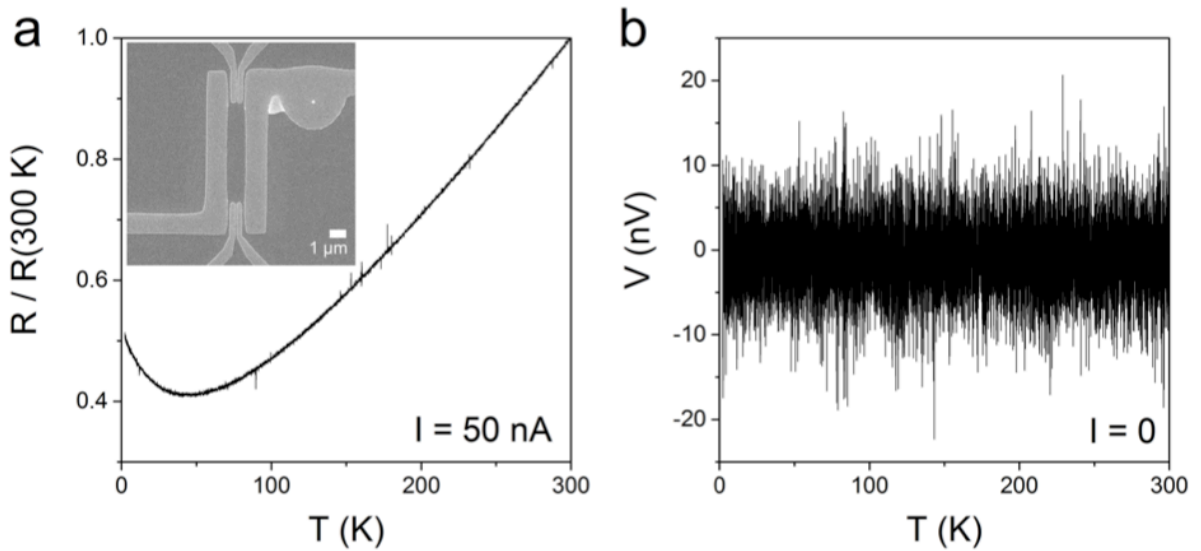

**Fig. S9:** (a) The temperature-dependent four-terminal resistance of the GdTO/STO/GdTO test structure with patterned closely-spaced electrodes, prepared without any etch processing. No enhanced low-temperature resistance fluctuations are observed at a measurement current of 50 nA (note that a few small spikes in the otherwise smooth data are measurement artifacts). The inset is an SEM image of the test structure, for which the large L-shaped electrodes at the left and right ends are the designed current electrodes, while the narrow closely spaced inner electrodes (~70 nm edge-to-edge inner separation) are the potential probes. The lack of enhanced low-T fluctuation level is confirmed in (b), which shows the measured potential difference between the potential probe contacts, without any applied current. Here, the potential fluctuates primarily within the level of ±10 nV over the entire temperature range, which is consistent with our measurement system noise floor.